\documentclass[12pt]{article}
\input epsf.tex

\renewcommand{\theequation}{\thesection.\arabic{equation}}
\newcommand{\be}{\begin{equation}}   \newcommand{\ee}{\end{equation}}
\newcommand{\bear}{\begin{eqnarray}}
\newcommand{\eear}{\end{eqnarray}}
\newcommand{\ba}{\begin{array}}      \newcommand{\ea}{\end{array}}
\newcommand{\lae}{\begin{array}{c}\,\sim\vspace{-21pt}\\< \end{array}}
\newcommand{\gae}{\begin{array}{c}\,\sim\vspace{-21pt}\\> \end{array}}
\begin{document}

\pagestyle{empty}
\begin{titlepage}
\def\thepage {}        % Kill page numbering

\title{\Large \bf
Electroweak Symmetry Breaking and \\ [2mm] Extra Dimensions  \\ [1cm]}

\author{\normalsize \bf  Hsin-Chia Cheng$^1$, Bogdan
A.~Dobrescu$^{2,3}$, and
Christopher T.~Hill$^{1,2}$ \\
\\
{\small {\it $^1$Enrico Fermi Institute }}\\
{\small {\it The University of Chicago }}\\
{\small {\it Chicago, Illinois 60637, USA \thanks{e-mail:
hcheng@theory.uchicago.edu, bdob@fnal.gov, hill@fnal.gov} }}\\
\\
{\small {\it $^2$Theoretical Physics Department}}\\
{\small {\it Fermi National Accelerator Laboratory}}\\
{\small {\it Batavia, Illinois 60510, USA }}\\
\\
{\small {\it $^3$Institute for Theoretical Physics}}\\
{\small {\it University of California}}\\
{\small {\it Santa Barbara, California 93106, USA}}\\
 }

\date{ }
\maketitle

   \vspace*{-14.9cm}
\noindent
\makebox[11.1cm][l]{\small hep-ph/9912343} {\small FERMILAB-PUB-99/358-T}
\\
\makebox[11.1cm][l]{December 14, 1999} {\small EFI-99-50} \\
%[1mm]
\makebox[11.1cm][l]{} {\small NSF-ITP-99-149} \\

 \vspace*{13cm}

\baselineskip=18pt

\begin{abstract}

   {
Electroweak symmetry can be naturally broken by
observed quark and gauge fields in various extra-dimensional
configurations. No new {\it fundamental}
fields are required below the quantum
gravitational scale ($\sim$ 10 -- 100 TeV).
We examine schemes in which the
QCD gauge group alone, in compact extra dimensions,
forms a composite Higgs doublet out of $(t,b)_L$ and a
linear combination of the Kaluza-Klein modes of $t_R$.
The effective theory at low energies is the Standard Model.
The top-quark mass is controlled by
the number of active $t_R$ Kaluza-Klein modes below the string scale,
and is in agreement with experiment.
}

\end{abstract}
\vfill
\end{titlepage}

\baselineskip=18pt
\pagestyle{plain}
\setcounter{page}{1}

%%%%%%%%%%%%%%%%%%%%%%%%%%%%%%%%%%%%%%%%%%%%%%%%%%%%%%%%%%%%
%%%%%%%%%%%%%%%%%%%%% Section 1 %%%%%%%%%%%%%%%%%%%%%%%%%%%%
%%%%%%%%%%%%%%%%%%%%%%%%%%%%%%%%%%%%%%%%%%%%%%%%%%%%%%%%%%%%

\section{Electroweak asymmetry and extra dimensions}
\setcounter{equation}{0}

There are two major experimental observations which are not explainable
solely in terms of the $SU(3)_C \times SU(2)_W \times U(1)_Y$ gauge
interactions and the three generations of quarks and leptons:
the electroweak symmetry breaking
and the existence of gravity. It is now widely believed that a quantum
theory of gravity necessitates a spacetime dimensionality greater than four.
In this paper we show that the extra spatial
dimensions, compactified at the $\sim$ TeV scale,
also provide simple and natural mechanisms for
electroweak symmetry breaking without the introduction of
explicit Higgs fields.

We will argue that the Standard Model is the effective
theory emerging, below the compactification scale, from a higher
dimensional $SU(3)_C \times SU(2)_W \times U(1)_Y$
gauge theory with three generations of quarks and leptons
and no fundamental Higgs field.
A composite Higgs
doublet arises naturally
in the presence of certain strongly coupled four-quark operators.
For concreteness, we will take these to involve typically
the left-handed top-bottom doublet ($\psi_L$) and a vector-like
quark \cite{seesaw}-\cite{minimal}, but we
anticipate many possible variations of this particular
arrangement. These particular four-quark operators
{\em are always
induced by QCD in compact dimensions}, via the exchange of the
Kaluza-Klein (KK) excitations of the gluons \cite{ewsb}.
Hence, the KK-gluons are effective ``colorons'' \cite{coloron} and 
their effects can be quite large because the higher-dimensional QCD 
coupling constant increases above the compactification scale.
The strength of these
contact interactions depends on the ratio of the compactification scale,
$M_c$, and the scale $M_s$ of the underlying quantum gravitational
effects. For $M_c$ in the TeV range \cite{tev, Carone:1999nz, Delgado:1999sv}, 
$M_s$ has to be around $10-100$ TeV such that
the quantum gravitational effects cut-off the non-renormalizable
higher-dimensional gauge interactions.
Hence, the measured weakness of the gravitational
interactions has to be explained by a modification of gravity at
short-distance, for instance as proposed in
refs.~\cite{largedim,Randall:1999ee,Lykken:1999nb}.

Indeed, the dependence of four-quark operator coefficients
on the $M_s/M_c$ ratio allows us to give a
nice connection with string/M theory  if one assumes that the
gauge couplings unify at the string scale \cite{Cheng:1999fu}. Due to
the power-law running of the gauge couplings in extra
dimensions \cite{dudas}, the value of the unified higher-dimensional
coupling, $g_{4+\delta}(M_s)$, and the $M_s/M_c$ ratio are determined
almost exclusively by the number $\delta$ of 
compact dimensions accessible to the Standard Model gauge bosons.
For $\delta \gae 2$, $g_{4+\delta}(M_s)$ is of order one in $M_s$ units,
corresponding to a string coupling of
order one. This is in accord with the argument based on dilaton stability
\cite{dine} that string theory is in the truly strong-coupling regime.
Furthermore, the large value of $g_{4+\delta}$ implies that the strength
of the four-quark operators induced by KK-gluon modes is
non-perturbative, and may indeed bind a composite Higgs.

The only remaining ingredient for a viable theory of dynamical
electroweak symmetry breaking is the above-mentioned vector-like quark.
In four dimensions, a composite Higgs doublet may be bound out of the
$\psi_L$ and the right-handed top field, $t_R$ \cite{bhl,Nambu}.
However, the Yukawa coupling of the Higgs doublet to its constituents is
typically large, so that the top quark mass
is too large (unless the theory is fine-tuned to 
nearly exact criticality, and the scale of the new
interactions is taken to the GUT scale; alternatively,
the measured top quark mass forces the VEV of this Higgs
doublet to be smaller than the Standard Model Higgs VEV, $v/\sqrt{2}$ where
$v \approx 246$ GeV is the electroweak scale).

On the other hand, if a new
vector-like fermion is introduced with the same
quantum numbers as $t_R$, it can then become the appropriate
constituent of the Higgs boson together with $\psi_L$.
The physical top mass is given in this case by a smaller eigenvalue
of a mass matrix involving the vector-like and top quarks
\cite{seesaw}. Therefore, such a seesaw mechanism
neatly accomodates both the measured top quark mass
and a Higgs VEV of $v/\sqrt{2}$.

It is quite striking that the Kaluza-Klein modes of the $t_R$
have exactly the quantum numbers of this requisite
vector-like quark. A key point of this paper is that the role of 
the vector-like quark can be naturally
played by the tower of KK modes of the $t_R$. Therefore, compact 
extra dimensions appear to provide everything needed for
a dynamical seesaw model of electroweak symmetry breaking\footnote{Other 
studies of electroweak symmetry breaking in extra
dimensions without a fundamental Higgs doublet can be found in 
\cite{more}.}.
Remarkably, however, while the vector-like
excitations are required, the seesaw mechanism
is no longer needed here, because
the top Yukawa coupling is automatically
suppressed by the (square-root of) number
of active KK modes of the $t_R$ with masses below $M_s$. Moreover,
for typical ratios of $M_s$ to the mass of the first quark KK
excitation, the top Yukawa coupling computed to leading order in 
$1/N_c$ is between $\sim 0.7$ and $\sim 1.4$.
Thus, the Standard Model value for the top Yukawa coupling 
($\sim 1$) is a natural consequence of our framework.

In Section 2 we first discuss chirality and anomaly cancellation
in the case of one extra dimension.
In Section 3 we present a detailed model of electroweak symmetry
breaking
valid below the quantum gravity scale which does not require any new
field
beyond the $SU(3)_C \times SU(2)_W \times U(1)_Y$ gauge fields and
the three generations of fermions, in a higher dimensional
configuration.
We study the low energy effects of this model in Section 4.
Finally, our conclusions are summarized in Section 5.

%%%%%%%%%%%%%%%%%%%%%%%%%%%%%%%%%%%%%%%%%%%%%%%%%%%%%%%%%%%%%%%%%%%%%%%%%%%%
%%%%%%%%%%%%%%%%%%%%% Section 2 %%%%%%%%%%%%%%%%%%%%%%%%%%%%%%%%%%%%%%%%
%%%%%%%%%%%%%%%%%%%%%%%%%%%%%%%%%%%%%%%%%%%%%%%%%%%%%%%%%%%%%%%%%%%%%%%%%%%%%%

\section{\hspace*{-4mm}Chirality and anomaly cancellation on a thick brane}
\setcounter{equation}{0}
%\hspace*{-2mm} 

In order to present the properties of the KK excitations of the $t_R$,
we start with a general discussion of fermions in five-dimensions.
The $t_R$ may be the zero-mode of a five-dimensional
fermion only if the gluons and hypercharge gauge boson propagate
in the fifth dimension. Therefore the extra dimension has to be
compact, with a radius below $\sim$ (3 TeV)$^{-1}$ \cite{
Carone:1999nz, Delgado:1999sv, Cheng:1999fu}.

%%%%%%%%%%%%%%%%%%%%%%%%%%%%%%%%%%%%%%%%%%%%%%%
\subsection{Chirality from orbifold projection}

A constraint on the compactification of the extra dimension comes
from the requirement that the $t_R$ is a chiral, two-component fermion.
The Lorentz group in five dimensions $SO(4,1)$ has only one
spin-1/2 representation which turns out to be non-chiral. The fermions
have four components, and the set of gamma matrices is formed of
the usual four-dimensional ones, $\gamma^\mu$, $\mu = 0,1,2,3$, and of
$i\gamma_5$.
Therefore, a chiral zero-mode of a five-dimensional
fermion may exist only if $SO(4,1)$ is broken. 
%A simple way of breaking $SO(4,1)$ while preserving $SO(3,1)$ is to 
%
This can be done by compactifying the fifth dimension on an orbifold,
or by imposing boundary conditions on the compact fifth dimension to
distinguish the left-
and right-handed components of the five-dimensional fermion.

Consider the four-dimensional Minkowski spacetime, with coordinates
$x^\mu$, and one additional transverse spatial dimension, with 
coordinate $y$. A simple way of
breaking $SO(4,1)$ while preserving the four-dimensional Lorentz 
invariance and allowing chiral fermions is to compactify
the fifth dimension, $y$, on an $S^1/{\bf Z}_2$ orbifold,
{\it i.e.}, a circle of radius $R=L/\pi$ with the identification 
$y \to -y$.
Five-dimensional fields are classified to be even or odd 
under ${\bf Z}_2$ parity. In terms of KK decomposition, 
the zero modes of the odd fields are projected out.
%and hence only even fields have zero modes.
The assignment of opposite ${\bf Z}_2$ parity to the left- and 
right-handed components of the five-dimensional fermion, 
$\chi_L(x, y)=-\chi_L(x, -y), \, \chi_R(x, y)
=\chi_R(x, -y)$, leaves massless only one
four-dimensional right-handed chiral fermion.

%There are no zero modes of $A_5$, which are scalar fields from the 
%four dimensional point of view. 

Equivalently, one may start by considering a five-dimensional
spacetime with boundaries along the fifth dimension 
at $y=0$ and $y = L$. A four-component fermion
field, $\chi(x,y)$, is defined on this space as a solution to the 
five-dimensional Dirac equation which obeys some conditions at 
$y = 0,L$.
%above spectra can be obtained by imposing some
%boundary conditions on the bulk fields at $y=0$ and $y=L$. 
The simplest chiral boundary conditions,
\bear
P_L \chi (x,0) & = & P_L \chi (x,L) = 0 ~,
\nonumber \\ [2mm]
\frac{\partial}{\partial y} P_R \chi (x,0) & = &
\frac{\partial}{\partial y} P_R \chi (x,L) = 0 ~,
\label{bc}
\eear
where $P_{L,R} = (1\mp \gamma_5)/2$,  lead to the 
quantization of momentum in the $y$ direction, and give rise to 
the same KK decomposition as the $S^1/{\bf Z}_2$ orbifold
projection discussed above. 

These boundary conditions may result
from the interactions between the bulk fields and the four-dimensional 
fields living on the branes located at 
$y=0, L$. This in fact could be
a physical explanation for the $S^1/{\bf Z}_2$ orbifold projection.
In this paper, however, we do not attempt to derive a theory 
valid at any energy scale, but rather we study an effective field theory 
in a compact higher-dimensional spacetime defined below an
ultra-violet cut-off
$M_s$. We therefore impose only {\it four}-dimensional
general covariance, and assume that the 
physics above $M_s$ does not generate unwanted operators.

The complete set of orthogonal functions on the $[0,L]$ interval
consistent with the boundary condition on $\chi_L$ (corresponding
to odd fields under $y\to -y$) is given by
\be
\sqrt{\frac{2}{L}}\, \sin\left(\frac{\pi j y}{L}\right) \, , \; j \ge 1
~.
\label{sine}
\ee
All these functions cancel on the boundaries, so that they indeed do not
include a zero-mode on the compact interval, $[0,L]$.
On the other hand, the boundary conditions for $\chi_R$ (corresponding
to even fields) allow a
complete set of orthogonal functions  on $[0,L]$,
%and for the even fields is given by
\be
\sqrt{\frac{1}{L}}\, ,  \;
\sqrt{\frac{2}{L}} \, \cos\left(\frac{\pi j y}{L}\right) \, , \; j \ge 1
~,
\ee
which includes a zero-mode.  % as for any even fields. 
The zero-mode of $\chi_R$ is identified as the right-handed top quark in
the weak eigenstate basis, $t_R$.
As a result, the decomposition of $\chi$ in KK modes is chiral:
\be
\chi (x,y) = \frac{1}{\sqrt{L}} \left\{ t_R(x) + \sqrt{2}
\sum_{j \ge 1} \left[ P_R \chi_R^j(x) \cos\left(\frac{\pi j y}{L}\right)
+ P_L \chi_L^j(x) \sin\left(\frac{\pi j y}{L}\right)  \right] \right\}
~.
\ee

A consequence of the boundary conditions (\ref{bc})
is that there is no fermion
mass term in the five-dimensional Lagrangian. Nevertheless, the Dirac
equation,
\be
\left( \gamma^\mu\partial_\mu + i \gamma_5 \frac{\partial}{\partial y}
\right) \chi (x,y) = 0 ~,
\ee
includes a $\gamma_5$ term so that it
cannot be decomposed in separate equations for the left-
and right-handed fermions.
It is straightforward to derive the fermion propagator for this
five-dimensional spacetime with the above boundary conditions:
\bear\!\!\!\!
\langle 0 | \,\chi(x^\prime,y^\prime)\, \overline{\chi}(x,y) \,| 0
\rangle
& = & \int \frac{d^4 k}{(2\pi)^4} e^{i k^\mu(x - x^\prime)_\mu}
\frac{2}{L}\sum_{j \ge 0} \left[\cos\left(\frac{\pi j
y^\prime}{L}\right) P_R
+ \sin\left(\frac{\pi j y^\prime}{L}\right) P_L\right]
\nonumber \\ [2mm]
& \times & \!\!\!\! \frac{\gamma^\mu k_\mu + \gamma_5 \pi j/L }
{k^\mu k_\mu - (\pi j/L)^2}
\left[ \sin\left(\frac{\pi j y}{L}\right) P_R
+\cos\left(\frac{\pi j y}{L}\right) P_L \right] 
\frac{i}{1 + \delta_{j0}}
\label{propagator}
\eear
We will use this propagator in section 3.2 to derive the Higgs 
potential.

%%%%%%%%%%%%%%%%%%%%%%%%%%%%%%%%%%%%%%%%
\subsection{Chiral Anomalies}

Next we study what happens when the $\chi$ fermion transforms under
some gauge symmetry. This is necessary in order to show that the
model of  electroweak symmetry breaking presented in the next section
is anomaly-free.

Under the $S^1/Z_2$ orbifold projection discussed in the previous
subsection, the ordinary four-dimensional spacetime components
of the gauge fields $A_\mu$ must be even while the fifth components 
$A_5$ must be odd, so that they have consistent interactions
with the $\chi$ fermion. Hence the fifth component $A_5$ has no zero
mode, and its KK modes become the longitudinal components of the 
heavy $A_\mu$ modes. 
The zero-mode of $A_5$ may also be eliminated
by imposing boundary conditions rather than an orbifold projection.
If the gauge fields propagate in the 
fifth dimension only on the $0\le y \le L$ interval, then the 
appropriate boundary conditions are given by
$A_5(x,0) = A_5(x,L) = 0$ and $\partial A_\mu (x,0)/ \partial y 
 = \partial A_\mu (x,L) / \partial y = 0$.
Although the graviton need not propagate at  $y > L$ or $y < 0$,
we refer loosely to the $[0,L]$ interval as a ``thick brane''
because $1/L$ is smaller than the string scale. 

The five-dimensional Lorentz-invariant gauge theories
have no chiral anomalies because the fermion representation is
vector-like.
However, the boundary conditions considered above prevent the existence
of a $\chi_L$ zero-mode, which raises the question of anomalies.
The $J^{a, r}_\chi \equiv \overline{\chi} \gamma^a T^r \chi$ current
has an anomaly given by
\be
D_a J^{a, r}_\chi = \frac{1}{24\pi^2 L}
\epsilon^{\mu\nu\lambda\rho}
{\rm Tr} \left[ T^r \partial_\mu \left( A_\nu \partial_\lambda A_\rho
+ \frac{1}{2} A_\nu A_\lambda A_\rho \right) \right]   ~,
\ee
where $A_\mu = -i g_5 A_{\mu}^{r^\prime} T^{r^\prime}$ is the gauge field,
the trace is over the products of group generators $T^r$,
$D^a$ is the covariant derivative,
and $\epsilon^{0123}=1$.
The index $a$ runs from 0 to 4, with
$\partial_4 \equiv \partial/\partial y$.
Throughout this paper we use latin (greek) indices to denote the
components of five (four) dimensional vectors.

Naively, one may think that this anomaly spoils the gauge invariance.
It turns out, however, that the anomaly in this five-dimensional theory
is more subtle.
This is because the action may include a Chern-Simons term on the
$[0,L]$ interval:
\be
{\cal L}_{\rm CS}(A) =
\frac{L-y}{96\pi^2 L}
\epsilon^{abcde} {\rm Tr} \left[ F_{ab} F_{cd} A_e
- \left( F_{ab} - \frac{2}{5} A_a A_b \right)A_c A_d A_e\right] ~,
\label{Chern}
\ee
where $F$ is the gauge field strength.
In the presence of the Chern-Simons term, the gauge current
becomes the sum of the fermion current and the Chern-Simons current.
As a result, the divergence of the total gauge current cancels
everywhere
on the {\it open} interval $(0,L)$:
\be
D_a \left(J^{a, r}_\chi +
J_{\rm CS}^{a, r} \right) = 0 ~.
\ee
Hence, the gauge theory with a Chern-Simons term
is well defined ({\it i.e.}, non-anomalous) in the bulk of the
fifth dimension.
This is to be contrasted with the
gauge anomaly in four-dimensions, which cannot be canceled by any
counterterm in the action.

The physical interpretation of anomaly cancellation in the bulk
of our five-dimensional theory is similar with that given in
ref.~\cite{Callan:1985sa} for the case of domain wall fermions
in 2+1 dimensions.
In the present case, the anomaly due to $t_R$ on the $[0,L]$ interval
produces gauge charges which
are collected by the Chern-Simons current and transported towards the
boundary. Therefore, in the bulk there is charge conservation.
At the boundary, though, the charges are lost, so that the five
dimensional theory with only one zero-mode fermion is indeed ill-behaved
due to the anomaly. This can be seen by computing the variation of the
action under a gauge transformation:
\be
\delta \int d^4 x \int^L_0 d y
\left( i \overline{\chi} \gamma^a D_a \chi + {\cal L}_{\rm CS} \right)
= \left.  L \int d^4 x \, D_a J^{a, r}_\chi \alpha^r \right|_{y = 0},
\label{action}
\ee
where $\alpha^r$ is the gauge transformation parameter.

Therefore, there is need for other fermions such that the overall anomaly
cancels, and the five-dimensional theory reduces to a non-anomalous
four-dimensional gauge theory at scales below $\pi/L$.
For simplicity we will assume that $t_R$ is the only fermion with
KK excitations below the string scale $M_s$. This is implemented in the 
effective field theory below $M_s$ by localizing all the Standard Model
fermions with the exception of $t_R$ at certain positions in the fifth
dimension. Evidently, the anomaly cancellation matches well in the
effective theory below the compactification
scale, where the only fermions present are the four-dimensional three
generations of quarks and leptons.

The microscopic implication of anomaly cancellation in this case is that
the charges which are driven by the Chern-Simons current (\ref{Chern})
to the boundary are brought by another Chern-Simons current to  the
location of the other third generation fermions where they are absorbed
by the corresponding four-dimensional anomalies.
For example, a left-handed fermion located at $y = y_0$ and $z=0$
requires a Chern-Simon term with a step function shape,
\be
\frac{\theta(y - y_0) - 1}{96\pi^2}
\, \epsilon^{abcde} {\rm Tr}
\left[ F_{ab} F_{cd} A_e
- \left( F_{ab} - \frac{2}{5} A_a A_b \right) A_c A_d A_e \right] ~,
\label{Chern-Simon}
\ee
to be added to ${\cal L}_{\rm CS}(A)$. As a result, the right-hand
side of eq.~(\ref{action}) vanishes and the theory is gauge invariant.

We emphasize that the five-dimensional gauge theory is 
non-renormalizable. The gauge coupling has mass dimension $(-1/2)$, and
it blows up at some scale $\sim M_s$. Therefore, any five-dimensional
gauge theory should be seen only as an effective field theory which
at the scale $M_s$ is replaced by a more fundamental framework,
such as string or M theory.
The Chern-Simons terms discussed here are supposed to be produced within
the theory that introduces the physical cut-off $M_s$.

Another possibility is that all third generation fermions are
defined on the $[0,L]$ interval with chiral boundary conditions similar
with those of $\chi$. In this case the overall Chern-Simons current 
vanishes and the anomalies are canceled exactly as in the four-dimensional 
Standard Model. However, this would imply that all third generation 
fermions have KK excitations, which would complicate the analysis of the 
model presented in the next section. To keep the discussion simple,
we will not investigate this possibility here.

%%%%%%%%%%%%%%%%%%%%%%%%%%%%%%%%%%%%%%%%%%%%%%%%%%%%%%%%%%%%%%%%%%%%%%%%%%%%%
%%%%%%%%%%%%%%%%%%%%% Section 3 %%%%%%%%%%%%%%%%%%%%%%%%%%%%%%%%%%%%%%%%%%%%
%%%%%%%%%%%%%%%%%%%%%%%%%%%%%%%%%%%%%%%%%%%%%%%%%%%%%%%%%%%%%%%%%%%%%%%%%%%%%%

\section{A model: right-handed top and QCD in extra
dimensions}
\setcounter{equation}{0}

In this Section we show that the dynamics in extra dimensions
allows us to construct a model of dynamical electroweak symmetry
breaking
without the need for a fundamental Higgs field.

Consider a $(4+\delta)$-dimensional spacetime with the four-dimensional
flat spacetime extended in the $x^\mu$, $\mu = 0,... ,3$ directions,
and extra spatial dimensions with coordinates $y$ and
$z_1, ... , z_{\delta-1}$.
Only some of the observed fields propagate in the extra dimensions.
The simplest configuration is that where
the gluons propagate in all these dimensions, the $t_R$ is the zero mode
of
a fermion, $\chi$, which  is fixed at $z = 0$ but propagates on
the $[0,L]$ interval in the $y$ dimension, and the
$\psi_L = (t,b)_L$ is fixed at $z = 0$ and $y = y_0$.
We choose $\delta \gae 3$ such that the effects of the gluons with momentum 
in the $z$ dimensions are non-perturbative when the $M_s$ scale is 
sufficiently large\footnote{Note that in the case of a single  
compact dimension, the four-quark operators induced by the tree level 
exchange of an infinite tower of gluon KK modes are finite.}.
We will assume that the gluons propagate on intervals of size $L$ and
$L_z$ in the $y$ and $z_1, ... , z_{\delta -1} $ dimensions,
respectively, with $L_z < L$.
In Fig.~1 we sketch the extra-dimensional configuration.

%%%%%%%%%%%%%%%%%%%%%%%%%%%%%%%%%%%%%%%%%%%%%%%%%%%%%%%%%%%
\begin{figure}[t]
\centering
\begin{picture}(200,100)(0,0)
%%%%%%
\put(20,10){\vector(0, 1){80}}
\put(-40,40){\vector(1, 0){300}}
\thicklines
\put(20,40){\line(1, 0){160}}
\put(20,41){\line(1, 0){160}}
\put(20,42){\line(1, 0){160}}
\put(20,60){\line(1, 0){160}}
\put(20,39.5){\line(1, 0){160}}
\put(20,40){\line(0, 1){20}}
\put(180,40){\line(0, 1){20}}
\put(50,37){\large $\times$} 
\put(25,90){$z_{1,...,\delta -1}$}
\put(240,30){$y$}
\put(9,30){\small $0$}
\put(175,30){\small $L$}
\put(5,60){\small $L_z$}
\put(50,30){\small $y_0$}
\end{picture}
\caption[]{
\label{Figure1}
\small The profile of the compact space. The $x$-coordinates of the flat
three-dimensional space
are transverse to the plane of the page. The $z_1, ..., z_{\delta - 1}$
coordinates  are depicted
collectively as one axis. The gluons propagate inside the rectangle,
the $ \chi$ propagates
along the $y$ axis, on the thick line, and the $\psi_L$ is located at
the point marked on the $y$ axis.}
\vspace{3mm}
\end{figure}
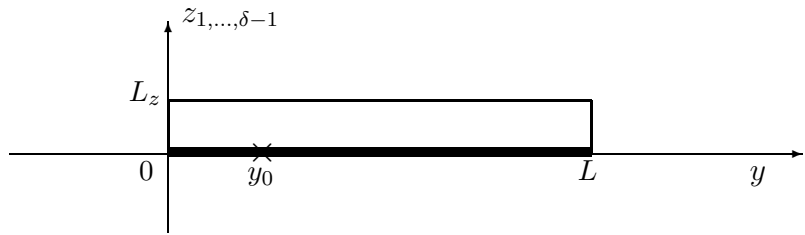
%%%%%%%%%%%%%%%%%%%%%%%%%%%%%%%%%%%%%%%%%%%%%%%%%%%%%%%%%%%

As mentioned in the previous section, it is convenient to assume that
all other quarks and the leptons are localized on four-dimensional
slices of $4+\delta$-dimensional spacetime, so that we do not have to
worry about their KK modes. Furthermore, if the left- and right-handed
fermions (other than $t_R$ and $\psi_L$) are split in the extra dimensions
\cite{Arkani-Hamed:1999dc}, then they cannot acquire large masses. Note
that this splitting does not produce the kind of flavor-changing neutral
currents discussed in \cite{Delgado:1999sv} provided the light fermions 
of same chirality are located at the same places.

The $U(1)_Y$ gauge bosons have to propagate in the $y$ dimension because
$\chi$ carries hypercharge.
The $SU(2)_W$ gauge bosons must propagate in the fifth dimension
only if different weak-doublet fermions are localized at different
places. Note that if gauge coupling unification is imposed, then it is
preferable to have the $SU(2)_W \times U(1)_Y$ gauge bosons propagating 
in the same space as the gluons.

%Finally, the graviton may propagate in a space of larger volume, which
%may or may not extend in the fifth dimension outside the $[0,L]$ interval.

%%%%%%%%%%%%%%%%%%%%%%%%%%%%%%%%%%%%%%%%%%%%%%%%%%%%%%%%%%%%%%%%%%%%%%%%%%%%
%%%%%%%%%%%%%%%%%%%%% SubSection 3.1 %%%%%%%%%%%%%%%%%%%%%%%%%%%%%%%%%%%%%%%
%%%%%%%%%%%%%%%%%%%%%%%%%%%%%%%%%%%%%%%%%%%%%%%%%%%%%%%%%%%%%%%%%%%%%%%%%%%%

\subsection{The five-dimensional theory}

After compactifying and integrating over the $z$ dimensions, we find a
tower of KK modes of the gluons, which are grouped in levels of
masses $\pi \sqrt{K}/L_z$ with $K$ a positive integer, and degeneracies
${\cal D}_K$
(${\cal D}_K = 0$ for some values of $K$, see Ref.~\cite{Cheng:1999fu}).
These gluon KK modes are five-dimensional fields whose effects at
energies
below their masses are approximately described by four-quark operators.

At scales between $\pi/L_z$ and the string scale, $M_s$, the dynamics
includes both light gluon KK modes and four-quark operators induced by
the heavier gluon KK modes. Although each gluon KK mode is weakly
coupled, the number of gluon KK modes may be sufficiently large 
\cite{ewsb} such that
the loop expansion breaks down. In order to analyze the effects of
this nonperturbative theory  below some scale $\Lambda$, we approximate
the dynamics of the gluons with momentum in the $z$ dimensions by
a five-dimensional effective theory with four-quark operators.
The matching between the five-dimensional low-energy theory and the
$(4+\delta)$-dimensional theory is likely to require the scale $\Lambda$
of the four-quark operators to be somewhere between $\pi/L_z$ and $M_s$.

By imposing that the loop expansion parameter \cite{Cheng:1999fu}
becomes of order one at $M_s$, we can estimate the separation between 
$\pi/L_z$ and $M_s$. For $\delta \gae 3$, the density of KK modes
is large and it turns out that $M_s$ is only about twice $\pi/L_z$.
Therefore, the uncertainty in $\Lambda$ is not worrisome.

The relevant piece of the five-dimensional Lagrangian density, involving
the four-dimensional $\psi_L (x^\mu)$ field and the five-dimensional
$\chi(x^\mu, y)$ and massless gluon fields is given at the scale
$\Lambda$ by
\be
{\cal L}_5 (x^\mu, y)  =
\delta(y - y_0) i \overline{\psi}_L  \gamma^\mu D_\mu \psi_L
+ \overline{\chi} \left(i\gamma^\mu D_\mu - \gamma_5 D_y\right) \chi
-\frac{1}{2 g_5^2} {\rm Tr}(F^{ab} F_{ab}) + {\cal L}_{\rm CS}(G) 
+ {\cal L}_{\rm int}
~.
\ee
$F^{ab}$ is the gluon field strength, ${\cal L}_{\rm CS}(G)$ is the
Chern-Simons term
for the gluon field [see eqs.~(\ref{Chern}) and (\ref{Chern-Simon})],
and $D$ is the covariant derivative:
\bear
D_\mu = \partial_\mu -  G_\mu ~,
\nonumber \\ [2mm]
D_y = \frac{\partial}{\partial y} -  G_y ~,
\eear
with $G_{\mu,y} = -i g_5 G_{\mu,y}^{r} T^{r}$ being 
five-dimensional gluon fields
(the zero modes from the KK expansion in the $z$ directions)
polarized in the $x^\mu$ and $y$ directions, respectively.
The five-dimensional strong coupling constant, $g_5$, has dimension
(mass)$^{-1/2}$.

The ${\cal L}_{\rm int}$ part of the ${\cal L}_5$ Lagrangian
includes the four-quark operators induced by
gluon KK mode exchange. Although the $SU(3)_C$ interactions are
flavor universal, the four-quark operators contained in 
${\cal L}_{\rm int}$
are not, because different quark fields are assumed to be localized at 
different positions in the extra dimensions. For example, all $SU(2)_W$
singlet quark fields other than $t_R$ and its excitations may be
localized at $z = z_R >0$, and the $SU(2)_W$ doublet quarks
of the first two generations may be localized at $z = z_L > 0$ with 
$z_L\neq z_R$. In this case the terms from ${\cal L}_{\rm int}$ that 
could lead to large quark masses in the low energy
theory, namely  the left-right current-current terms, are exponentially
suppressed unless they involve only $\psi_L$ and $\chi$.

The four-quark operators involving $\psi_L(x^\mu)$ and $\chi(x^\mu, y)$,
obtained by integrating out the five-dimensional gluon KK excitations,
are given by
\be
{\cal L}_{\rm int} (x, y)  = - \frac{cg_5^2}{2\Lambda^2} \left\{ \left[
\delta(y - y_0) \left(\overline{\psi}_L \gamma^\mu T^r \psi_L\right)
+ \left(\overline{\chi} \gamma^\mu T^r \chi\right) \right]^2
+ \left(\overline{\chi} \gamma_5 T^r \chi\right)^2 \right\} ~,
\ee
where $c \gg 1$ is a dimensionless coefficient obtained by summing over
the effects of the gluon KK modes, and $T^r$ are $SU(3)_C$ generators.

These four-quark operators may be Fierz transformed, with the result
\be
{\cal L}_{\rm int} (x, y) = \frac{cg_5^2}{\Lambda^2}
\left\{ \delta(y - y_0)
\left(\overline{\psi}_L \chi\right) \left(\overline{\chi} \psi_L\right)
+ \frac{5}{16}\left[\left(\overline{\chi}\chi\right)^2
- \frac{1}{3} \left(\overline{\chi}\gamma_5\chi\right)^2 \right]
\right\}
+ ... ~,
\label{four}
\ee
where the ellipsis stand for vectorial and tensorial four-quark
operators, which are irrelevant at low energies.

%%%%%%%%%%%%%%%%%%%%%%%%%%%%%%%%%%%%%%%%%%%%%%%%%%%%%%%%%%%%%%%%%%%%%%%%%%%%
%%%%%%%%%%%%%%%%%%%%% SubSection 3.2 %%%%%%%%%%%%%%%%%%%%%%%%%%%%%%%%%%%%%%%
%%%%%%%%%%%%%%%%%%%%%%%%%%%%%%%%%%%%%%%%%%%%%%%%%%%%%%%%%%%%%%%%%%%%%%%%%%%%

\subsection{The five-dimensional effective potential}

The operators shown in ${\cal L}_{\rm int}$ provide attractive
interactions which give rise to bound states: a four-dimensional
weak-doublet complex scalar, $H(x^\mu)$, and a five-dimensional
gauge singlet real scalar, $\varphi(x^\mu, y)$.
These composite scalars are propagating degrees of freedom only below
the compositeness scale. According to our approximation in which the
full KK mode dynamics is described at low energy by a
five-dimensional theory with four-quark operators,
the compositeness scale is identified with $\Lambda$.

At the compositeness scale the composite scalars are non-propagating,
and the four-quark operators may be replaced by
Yukawa interactions between the scalars and their constituents.
The first two terms shown in (\ref{four}) are equivalent with
\be
{\cal L}_c[\Lambda] = - \delta(y - y_0) \left[
\sqrt{ c g_5^2} \left(\overline{\chi} \psi_L\right) H
+ \Lambda^2 H^\dagger H \right] 
- \sqrt{\frac{5}{8} c g_5^2} \left(\overline{\chi}\chi\right) \varphi
- \frac{\Lambda^2}{2} \varphi^2 ~,
\ee
as can be seen by integrating out $H(x^\mu)$ and $\varphi(x^\mu, y)$.
The last term in eq.~(\ref{four}) gives rise to a five-dimensional
pseudo-scalar. However, the coefficient of this term is suppressed
by the factor of 1/3, such that the pseudo-scalar is not sufficiently
deeply-bound to be relevant at energies below the compositeness scale.

At scales $\mu < \Lambda$, the Yukawa interactions induce kinetic
terms for the scalars: 
\bear
{\cal L}_c[\mu] & = & \delta(y - y_0) \left[ Z_H(\mu) D^\nu H^\dagger
D_\nu H - \sqrt{ c g_5^2} \left(\overline{\psi}_L \chi\right) H\right]
\nonumber \\ [2mm]
&& + Z_\varphi(\mu) \partial^a \varphi \partial_a \varphi
- \sqrt{\frac{5}{8} c g_5^2} \left(\overline{\chi}\chi\right) \varphi
- V(\mu) ~.
\label{kinetic}
\eear
%%%%%%%%%%%%%%%%%%%%%%%%%%%%%%%%%%%%%%%%%%%%%%%%%%%%%%%%%%%
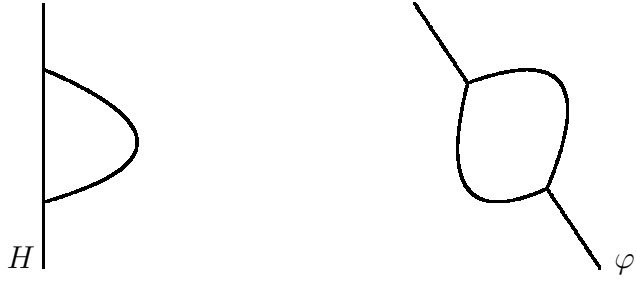
\begin{figure}[t]
\begin{picture}(388,85)(-50,30)
%%%%%%
\thicklines
\put(80,50){\line(0, 2){100}}
\qbezier(80,125)(150,95)(80,75)
\multiput(240,120)(160,0){1}{\qbezier(0,0)(-10,15)(-20,30)}
\multiput(270,80)(160,0){1}{\qbezier(0,0)(10,-15)(20,-30)}
\qbezier(240,120)(295,140)(270,80)
\qbezier(270,80)(225,60)(240,120)
\put(66,50){$H$}
\put(296,50){$\varphi$}
\end{picture}
\caption[]{
\label{Figure2}
\small  Large-$N_c$ contributions to the composite scalar self-energies.
The vertical lines are four dimensional fields localized at $y = y_0$,
and the curved or slanted lines are five-dimensional fields.
The external lines represent the $H$ and $\varphi$, while in the loops 
run the $\psi_L$ and $\chi$ quarks.\\ }
\end{figure}
%%%%%%%%%%%%%%%%%%%%%%%%%%%%%%%%%%%%%%%%%%%%%%%%%%%%%%%%%%%
The wave function renormalization $Z_H$ can be determined by computing
the self-energy of the weak-doublet in the large-$N_c$ limit 
(see Fig.~2):
\be
Z_H(\mu) = 2 N_c \frac{c g_5^2}{L}
\sum_{j \ge 0} \frac{\cos^2(\pi j y_0/L)}{\left(1+\delta_{j0}\right)}
\int \frac{d^4 k}{(2\pi)^4} \,
\frac{-i}{k^\mu k_\mu \left[ k^\nu k_\nu - (\pi j/L)^2 \right]}  ~.
\ee
The integral is logarithmic divergent, and has to be cut-off at
$\Lambda$.
The sum over the momenta in the fifth dimension is also divergent,
and is cut-off at $n_{\rm KK} \approx \Lambda L/\pi$.
The integral has also an infrared cut-off at $\mu$.

The wave function renormalization for the $\varphi$ scalar has a more
complicated form, due to the two $\chi$ propagators involved
[see eq.~(\ref{propagator})]. Keeping only the leading divergent piece,
we find
\be
Z_\varphi(\mu) \approx \frac{5}{4} N_c \frac{c g_5^2}{L}
\sum_{j \ge 0} \int \frac{d^4 k}{(2\pi)^4}\,
\frac{-i}{\left[ k^\nu k_\nu - (\pi j/L)^2 \right]^2 }  ~.
\ee
Note that the wave function renormalization for
$\partial \varphi/\partial y$ is somewhat arbitrary (it can be absorbed in 
the mass term for $\varphi$), and does not have to be the same as 
$Z_\varphi(\mu)$.  In ${\cal L}_c[\mu]$ we have chosen these two
wave function renormalizations to be the same for convenience.

The scalar potential includes mass and quartic terms,
\be
V(\mu) = \delta(y - y_0) \left[ \frac{\tilde{\lambda}_H}{2}
\left(H^\dagger H\right)^2
+ \frac{\tilde{\lambda}_0L}{2} H^\dagger H \varphi^2
+ \tilde{M}_H^2 H^\dagger H \right]
+ \frac{\tilde{\lambda}_\varphi L}{4!}\varphi^4
+ \frac{\tilde{M}_\varphi^2}{2} \varphi^2 ~,
\label{potential}
\ee
as well as higher-dimensional terms which we will ignore.
The mass parameters computed in the large-$N_c$ limit are given by
\bear
\tilde{M}_H^2(\mu) & = & \Lambda^2 - 4 N_c \frac{c g_5^2}{L}
\sum_{j \ge 0} \frac{\cos^2(\pi j y_0/L)}{1+\delta_{j0}}
\int \frac{d^4 k}{(2\pi)^4}\, \frac{i}{ k^\nu k_\nu - (\pi j/L)^2 } ~,
\nonumber \\ [2mm]
\tilde{M}_\varphi^2(\mu) & \approx & \Lambda^2
- \frac{5}{2} N_c \frac{c g_5^2}{L} \sum_{j \ge 0}
\int \frac{d^4 k}{(2\pi)^4} \, \frac{i}{ k^\nu k_\nu - (\pi j/L)^2} ~.
\label{mass}
\eear
In the expression for $\tilde{M}_\varphi^2$ we have kept again only the
leading divergent piece.
%%%%%%%%%%%%%%%%%%%%%%%%%%%%%%%%%%%%%%%%%%%%%%%%%%%%%%%%%%%
\begin{figure}[t]
\begin{picture}(388,110)(60,20)
%%%%%%
\thicklines
\put(80,50){\line(0, 2){100}}
\multiput(0,0)(160,0){3}{
  \qbezier(80,80)(110,60)(160,80)
  \qbezier(80,120)(130,140)(160,120) } 
\multiput(161,50)(160,0){2}{\line(0, 2){100}}
\qbezier(480,120)(490,135)(500,150)
\qbezier(480,80)(490,65)(500,50)
\multiput(240,120)(160,0){2}{\qbezier(0,0)(-10,15)(-20,30)}
\multiput(240,80)(160,0){2}{\qbezier(0,0)(-10,-15)(-20,-30)}
\qbezier(480,80)(500,95)(480,120)
\multiput(160,0)(160,0){2}{ \qbezier(80,80)(60,105)(80,120)}
\end{picture}
\caption[]{
\label{Figure3}
\small  Large-$N_c$ contributions to the $\tilde{\lambda}_H,
\tilde{\lambda}_0$ and $\tilde{\lambda}_\varphi$ quartic couplings.
The lines represent fields as explained in the caption of Fig.~2.\\ }
\end{figure}
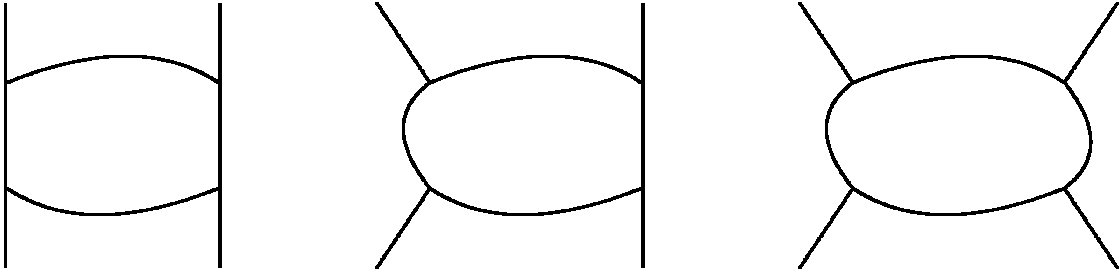
%%%%%%%%%%%%%%%%%%%%%%%%%%%%%%%%%%%%%%%%%%%%%%%%%%%%%%%%%%%

In the large-$N_c$ limit, the  leading contribution to the
dimensionless quartic coupling, $\tilde{\lambda}_H$,
is given by a quark loop with alternating $\chi$ and $\psi_L$
propagators (Fig.~3). Therefore, the result can be written as a double sum 
over the $\chi$ momenta in the fifth dimension:
\be
\tilde{\lambda}_H(\mu) = 8 N_c \left(\frac{c g_5^2}{L}\right)^{\! 2}
\sum_{j_{1,2} \ge 0} f_{j_1 j_2} \cos^2\left(\frac{\pi j_1 y_0}{L}
\right) \,\cos^2\left(\frac{\pi j_2 y_0}{L}\right)
\ee
where we have defined
\be
f_{j_1 j_2} \equiv \frac{1}
{\left(1+\delta_{j_1 0}\right) \left(1+\delta_{j_2 0}\right)}
\int \frac{d^4 k}{(2\pi)^4}
\frac{-i}{ \left[k^\nu k_\nu - (\pi j_1/L)^2 \right]
\left[k^\rho k_\rho - (\pi j_2/L)^2\right]} ~.
\ee
The coefficients of the quartic terms involving $\varphi$ have mass
dimension $-1$. The factors of $L$ are introduced in eq.~(\ref{potential})
such that the $\tilde{\lambda}_0$ and $\tilde{\lambda}_\varphi$
quartic couplings are dimensionless:
\bear
\tilde{\lambda}_0(\mu) & \approx & \frac{5}{4} N_c
\left(\frac{c g_5^2}{L}\right)^{\! 2}
\sum_{j_{1,2,3} \ge 0} f_{j_1 j_3}
\frac{d_{j_1 j_2} d_{j_3 j_2} }{\left(1+\delta_{j_2 0}\right)}
\cos\left(\frac{\pi j_1 y_0}{L}\right)
\,\cos\left(\frac{\pi j_3 y_0}{L}\right)
 \nonumber \\ [2mm]
\tilde{\lambda}_\varphi(\mu) & \approx & \frac{75}{128} N_c
\left(\frac{c g_5^2}{L}\right)^{\! 2}
\sum_{j_{1,2,3,4} \ge 0} f_{j_1 j_2}
\frac{d_{j_1 j_2} d_{j_3 j_2}d_{j_3 j_4} d_{j_1 j_4}}
{\left(1+\delta_{j_3  0}\right)
\left(1+\delta_{j_4 0}\right)} ~.
\eear
When all the $\varphi$ fields from the $\varphi^4$ interaction 
have momentum $\pi/L$ in the $y$ direction, we obtain:
\be
d_{j_1 j_2} \equiv
\delta_{j_2, j_1+1} - \delta_{j_2, j_1-1} + \delta_{j_2, 1-j_1} ~.
\label{coefficient}
\ee
%%%%%%%%%%%%%%%%%%%%%%%%%%%%%%%%%%%%%%%%%%%%%%%%%%%%%%%%%%%%%%%%
To evaluate all these parameters, we assume that the number of the
KK modes in the $y$ direction, $n_{\rm KK}$, is large enough so that
we can approximate the sums over KK states by integrals.
The expressions obtained for the parameters are given in the Appendix.

The kinetic terms in ${\cal L}_c[\mu]$ [see eq.~(\ref{kinetic})]
may be cannonically normalized
by redefining the scalar fields: $H \rightarrow H\sqrt{Z_H}$ and
$\varphi \rightarrow \varphi\sqrt{Z_\varphi}$.
In this case, the terms in the scalar potential
have the same form as in eq.~(\ref{potential}), but with appropriately
normalized coefficients. We denote the new parameters by dropping
the tilde from the corresponding symbols used in eq.~(\ref{potential}).
The squared-masses are given by
\bear
M_H^2 & = & \frac{\tilde{M}^2_H}{Z_H} \approx \frac{2 \Lambda^2}{F_1(y_0)}
\left[\frac{4 \pi^2}{n_{\rm KK} N_c c g_s^2 } - F_3(y_0) \right]
\nonumber \\ [3mm]
M_\varphi^2 & = & \frac{\tilde{M}^2_\varphi}{Z_\varphi} \approx
\frac{2 \Lambda^2}{F_2}
\left(\frac{32 \pi^2}{5 n_{\rm KK} N_c c g_s^2 } - F_4 \right)
\label{masses}
\eear
We have used here the four-dimensional $SU(3)_C$ gauge coupling, $g_s$,
obtained in terms of the five-dimensional coupling 
by integrating over the $y$ dimension:
\be
g_s = \frac{g_5}{\sqrt{L}} ~.
\ee 
The dependence of $M_H^2$ on the position $y_0$ of the $\psi_L$ doublet
is encoded in the $F_{1,3}(y_0)$ functions, which are symmetrical
under the $y_0 \rightarrow L-y_0$ reflection.
 $F_1(y_0)$ and $F_3(y_0)$ have values of order one, with maxima on 
the boundary and minima at $y = L/2$. $F_2$ and $F_4$ are constant functions
on the $[0,L]$ interval because the $\varphi$ mass is 
induced by interactions which conserve momentum in the $y$ dimension.
These functions are given in terms of 
divergent sums and integrals and depend on the cut-off procedure.
In the Appendix we estimate them in the continuum limit with a specific 
cut-off. 

Similarly, the quartic couplings may be written as follows:
\bear
\lambda_H & = & \frac{\tilde{\lambda}_H}{Z_H^2} \approx 
\frac{32\pi^2 F_5(y_0)}{N_c \left[F_1(y_0)\right]^2} ~,
\nonumber \\ [3mm]
\lambda_0 & = &
\frac{\tilde{\lambda}_\varphi}{3 Z_H Z_\varphi} 
\approx  \lambda_\varphi \frac{F_6(y_0)}{F_1(y_0)} ~,
\nonumber \\ [3mm]
\lambda_\varphi & = & \frac{\tilde{\lambda}_\varphi}{Z_\varphi^2}
\approx \frac{16\pi^2}{ n_{\rm KK} N_c F_2} ~.
\label{lambda}
\eear
Like the other $F$-functions written in the Appendix, $F_{5,6}(y_0) \sim 1$. 
Note that $\lambda_H$ is enhanced by an $n_{\rm KK}$ factor compared with
the other quartic couplings.

%%%%%%%%%%%%%%%%%%%%%%%%%%%%%%%%%%%%%%%%%%%%%%%%%%%%%%%%%%%%%%%%%%%%%%%%%
%%%%%%%%%%%%%%%%%%%%% Section 4 %%%%%%%%%%%%%%%%%%%%%%%%%%%%%%%%%%%%%%%%%%%%
%%%%%%%%%%%%%%%%%%%%%%%%%%%%%%%%%%%%%%%%%%%%%%%%%%%%%%%%%%%%%%%%%%%%%%%%%%%%%%

\section{Four-dimensional phenomenology}
\setcounter{equation}{0}

The squared-mass parameters from the five-dimensional potential may turn
negative if the four-quark operators induced by gluon KK modes are strong
enough. Therefore, the four-dimensional field, $H$, and the
five-dimensional real scalar, $\varphi$, may acquire VEVs.
The composite weak-doublet $H$ may be identified with the
Standard Model Higgs doublet.
In this Section we discuss  the scalar spectrum and its phenomenological 
implications, and we estimate the top quark mass.

%%%%%%%%%%%%%%%%%%%%%%%%%%%%%%%%%%%%%%%%%%%%%%%%%%%%%%%%%%%%%%%%%%%%%%%%%%%
\subsection{Higgs boson mass}

First, we consider the case in which $\psi_L$ is 
located at the boundary ($y_0=0$).
An inspection of the squared-masses computed in the 
large-$N_c$ limit [see eq.~(\ref{masses})] reveals that
only $M_H^2$ should become negative because the coupling
in the $\overline{\chi} \psi H$ channel is stronger than the coupling in
the $\overline{\chi} \chi \varphi$ channel. 
In addition, the four-dimensional quartic coupling involving 
both $H$ and $\varphi$ vanishes in this case because the $\varphi$
has a zero wave function on the boundary.
This implies that there is no mixing between $H$ and $\varphi$.
Therefore, the $\varphi$ has no effect on the Higgs potential in this case.
The $H$ acquires a VEV while the KK modes of $\varphi$ have masses
above the compactification scale.

The effective theory below the compactification scale is given by the 
Standard Model.
The compositeness of the Higgs doublet is not manifest at low-energy.
However, as a remnant of the strong dynamics that binds the Higgs,
the quartic Higgs coupling is large, $\lambda_H \gg 1$.
The Higgs boson mass $M_{h^0}$, given at tree level by 
$v \sqrt{\lambda_H (v) }$,
appears to be above 1 TeV. The tree level estimate, though, should not 
be taken too seriously due to the large $\lambda_H$.
Because the theory that gives rise to the composite Higgs boson is 
unitary (above the $M_s$ scale, the unitarity should be enforced by quantum 
gravitational effects), the Higgs mass is below the  bound imposed by 
the unitarity of the $WW$ scattering cross section in the Standard Model.
Once the non-perturbative 
corrections to $M_{h^0}$ are included, we expect $M_{h^0} \sim {\cal O}(1/2)$
TeV. Generically, when the $\psi_L$ is at $y_0=0$, the Higgs boson is a broad 
resonance.

Note that such a heavy Higgs boson is perfectly compatible with 
the electroweak precision data. The often quoted upper bound 
on the Higgs boson based on the fit to the electroweak data is 
valid only if there are no fields or interactions beyond the 
Standard Model \cite{Higgs}.
In our case, however, there are KK excitations of the 
Standard Model gauge bosons and $t_R$, with masses in the TeV range.
In their presence, a heavier Higgs  boson is not only allowed, but potentially
preferred by the fit to the data.
This has been shown in the context of extra dimensions in ref.~\cite{wells}.
Specifically, the shift in the electroweak observables due to the 
mixing of the $W$ and $Z$ with their KK excitations 
reproduces that due to a light Higgs boson (when the Higgs is trapped 
on a 3+1-dimensional wall, like in our case).
Furthermore, when a vector-like quark identical with our KK modes of $t_R$
is added to the Standard Model, the fit to the electroweak data yields a 
heavy Higgs for a vector-like quark mass around 5 TeV \cite{new}.
Of course, when the vector-like quark is much heavier, or 
equivalently the compactification scale in our model is increased, 
one recovers the Standard Model in the decoupling limit. Therefore, the 
$y_0=0$ case is consistent  with
the electroweak precision data only if the compactification scale 
is not above ${\cal O}(10$ TeV).

In the other case, where the $\psi_L$ fermion doublet
is located in the middle of the interval occupied by
$\chi$, {\it i.e.} $y_0 \sim L/2$, both $H$ and $\varphi$ 
may develop VEVs. (Note that eqs.~(\ref{masses}) imply that
for $F_3(y_0) \approx 5F_4/8 $
both $M_H^2$ and $M_\varphi^2$ turn negative at some particular 
value of $n_{\rm KK}\,cg_s^2$.)
Since the Higgs VEV is below the compactification scale, it is
appropriate to integrate first over the fifth dimension, and only 
afterwards to minimize the potential.
The five-dimensional real scalar decomposes in a tower of KK modes
\be
\varphi(x^\mu, y) = \sqrt{\frac{2}{L}} \sum_{j \ge 1} \varphi_j(x^\mu)
\sin\left(\frac{\pi j y}{L}\right)  ~.
\ee
It is likely that only one or the first few modes of $\varphi$ are
light enough to have a significant mixing
with the $H$. 

For simplicity, we consider that the Higgs field mixes with
one $\varphi$ mode.
The four-dimensional potential may be obtained readily from 
eq.~(\ref{potential}):
\be
V_4 = \frac{\lambda_H}{2} \left(H^\dagger H\right)^2
+ \lambda_0(y_0) \sin^2\!\left(\frac{\pi y_0}{L}\right) 
H^\dagger H \varphi_1^2
+ \frac{\lambda_\varphi}{16}\varphi_1^4
+ M_H^2 H^\dagger H 
+ \frac{1}{2} \left(M_\varphi^2 + \frac{\pi^2}{L^2} \right)\varphi_1^2 
~.
\label{potential-4}
\ee
After the scalar potential is minimized and the
scalar fields shifted, we find the following mass terms for
the two light neutral degrees of freedom:
\be
\frac{1}{2} \left( h, \phi \right)
\left( \begin{array}{cc} \lambda_H v^2 & 
2\lambda_0(y_0) v u \sin^2\!\left(\frac{\pi y_0}{L}\right)\\ [3mm]
2\lambda_0(y_0) v u \sin^2\!\left(\frac{\pi y_0}{L}\right)
&  \frac{1}{2} \lambda_\varphi u^2
\end{array} \right) \left( \begin{array}{c} h \\ [3mm] \phi
\end{array}\right)~,
\label{matrix}
\ee
where $v \approx 246$ GeV and $u$ is the $\varphi_1$ VEV.

In the limit $\lambda_H v^2 \ll \lambda_\varphi u^2$,
the mixing of $h$ and $\phi$ decreases the Higgs boson mass:
\be
M_{h^0}^2 \approx \lambda_H v^2 
\left[1 - \frac{8[\lambda_0(y_0)]^2}{\lambda_H\lambda_\varphi}
\sin^4\!\left(\frac{\pi y_0}{L}\right)\right]
\ee
For $y_0 = L/2$, the $M_{h^0}$ decreases by $\sim (70/n_{\rm KK})\%$. This 
value is derived using the $\lambda_H$ given in eq.~(\ref{lambda}). 
As argued before, we expect that the quantum corrections
actually drive $\lambda_H$ smaller, which would lead to an enhancement
of the change in $M_{h^0}$ due to mixing. 
If more $\varphi$ modes participate in the mixing, the decrease 
in $M_{h^0}$ becomes even more significant. 
Perhaps the Higgs boson may be driven close to the current LEP limit.
Unfortunately, it is hard to study
the scalar spectrum in general, with all KK modes included, especially
given that the parameters of the full effective potential are 
not accurately known.

In the other limit, where $\lambda_H v^2 \gg \lambda_\varphi u^2$,
the $h-\phi$ mixing may be ignored. The Higgs boson remains heavy,
but the physical $\phi^0$ scalar could be very light. Its mass,
\be
M_\phi^0 \approx u \sqrt{\frac{\lambda_\varphi}{2}} ~, 
\ee
is not constrained by the consistency of the model. The experimental 
lower bounds on a neutral scalar which couples only to the top quark
and the Higgs boson are quite weak \cite{minimal}.
It is therefore possible that the  Higgs boson decays into $\phi^0$
pairs, giving rise to unusual signals at future collider experiments
\cite{light}.

%%%%%%%%%%%%%%%%%%%%%%%%%%%%%%%%%%%%%%%%%%%%%%%%%%%%%%%%%%%%%%%%%%%%%%
\subsection{Top quark mass prediction}

We can now predict the top-quark mass as a function of 
the number of KK modes and the position $y_0$ of the $\psi_L$ doublet.
The fermion couplings to the composite scalars are given 
by eq.~(\ref{kinetic}). Upon normalization of the scalar kinetic 
terms and integration over the $y$ dimension, the Yukawa couplings become:
\bear
&& - \xi_t \sum_{j = 1}^{n_{\rm KK}} 
\left(\frac{2}{1 + \delta_{j0}}\right)^{\! 1/2}
\cos\!\left(\frac{\pi j y_0}{L}\right)
\overline{\chi}_R^j \psi_L H 
\nonumber \\ [2mm]
&& -  \xi_\chi \sum_{j_{1,2,3} = 1}^{n_{\rm KK}} 
\left(\delta_{j_3, j_1+j_2} - \delta_{j_3, j_1-j_2} + \delta_{j_3, j_2-j_1}
\right) \overline{\chi}_L^{j_1} \chi_R^{j_2}\varphi_{j_3} 
 + {\rm h.c.}
\label{yukawa}
\eear
Note that the Yukawa couplings of the Higgs doublet depend on the position
in the fifth dimension.
The zero-mode of $\chi$, namely $t_R$, has a Yukawa coupling to $H$ given by
\be
\xi_t = \frac{2\sqrt{2}\pi }{\sqrt{N_c n_{\rm KK} F_1(y_0)}} ~.
\ee
The Yukawa couplings of the $\varphi$ KK modes are position-independent 
due to momentum conservation at the $\overline{\chi}\chi \phi$ vertex:
\be
\xi_\chi = \frac{2\pi }{\sqrt{N_c n_{\rm KK} F_2}} ~.
\ee

The fermion masses for the $t_L$ component of $\psi_L$ and the 
KK modes of $\chi$ form a $(n_{\rm KK}+1)\times (n_{\rm KK}+1)$
matrix. There are two contributions to the elements of this mass
matrix. First, the Yukawa interactions give contributions
determined by replacing the $H$ and $\varphi^j$ scalars with their VEVs
in eq.~(\ref{yukawa}).
Second, the kinetic term of the five-dimensional $\chi$ field 
yields the usual KK mass terms: 
\be
\sum_{j = 1}^{n_{\rm KK}} \frac{\pi j}{L} \overline{\chi}_L^{j} 
\chi_R^{j} ~.
\ee

In the case where $y_0 = 0$, the fermion mass matrix is easy to write:
\be
\left( \overline{t}_L \, , \; \overline{\chi}_L^1  \, , \; 
\overline{\chi}_L^2  \, , \; ... \right)
\left( \begin{array}{ccccc}
\frac{\textstyle \xi_t v}{\textstyle \sqrt{2}}  & \xi_t v  & 
\xi_t v  & ... 
\\ [3mm]
0 & \frac{\textstyle \pi}{\textstyle L} & 0 & ... \\ [3mm]
0 & 0 & \frac{\textstyle 2\pi}{\textstyle L} &... \\ [3mm]
...  & ... & ... & ... \end{array} \right)
\left( \begin{array}{c}  t_R  \\ [3mm] \chi_R^1
\\ [3mm] \chi_R^2, \\ [3mm] ... \end{array} \right)
+ {\rm h.c.} 
\ee
The top-quark mass (in the limit where we ignore the small
mixing of the top with the charm and up) is given by the 
lowest eigenvalue of the above mass matrix.
It is amusing that this matrix has the same form as the 
one for neutrino masses given in ref.~\cite{Arkani-Hamed:1998vp}.
Note that our assumption that the KK-gluons in the
$z$ dimensions may be integrated out below the cut-off scale $\Lambda$
(see Section 3) is legitimate provided $\Lambda \gg \pi/L$.
Thus, to be consistent we must impose $n_{\rm KK} \gae 10$.
Expanding in $(vL/\pi)^2 \ll 1$, and taking $n_{\rm KK} \gg 1$, we find
\be
m_t \approx \frac{\xi_t v}{\sqrt{2}} 
\left[1 - \frac{3}{2} \left(\xi_t v L\right)^2 \right] ~.
\ee
For $L \lae 1$ TeV$^{-1}$, the second term gives a small correction 
$(\!\lae 1/n_{\rm KK})$ to $m_t$.
Therefore, the top mass is predicted in terms of $n_{\rm KK}$:
\be
m_t \approx \frac{600 \; {\rm GeV}}{\sqrt{n_{\rm KK}}} ~.
\ee
The measured top mass can be used to determine the 
number of top KK modes: 
\be
n_{\rm KK} \approx 12.
\ee
The number of top KK modes is related to the cut-off scale
$\Lambda \approx n_{\rm KK} \pi/L$, which is of the order of the 
string scale $M_s$.
If the first KK modes have a mass of a few TeV, then the above 
prediction determines the scale of quantum gravity $M_s \sim 30$ TeV.

Furthermore, given that a cut-off scale significantly above 
$\sim 50$ TeV
would require excessive fine-tuning (we assume that the theory is not 
supersymmetric below the string scale), we find a naturalness upper bound
 $n_{\rm KK} \lae 20$. Therefore, instead of using the measured top mass
to determine the number of KK modes, we may reverse the argument and 
determine the typical values of the top mass in our model.
For $10 \lae  n_{\rm KK} \lae 20$, we find a range, 
$130 \; {\rm GeV} \lae m_t  \lae 190 \; {\rm GeV}$, which within 
the theoretical uncertainties is in agreement with the measured value. 

When the $\psi_L$ is placed in the middle of the thick brane occupied 
by $\chi$, {\it i.e.} $y_0 \sim L/2$, some of the $\varphi$ KK modes
may acquire VEVs, as discussed in section 4.1. Therefore, the fermion
mass matrix becomes more complicated to analyze. If only the first 
$\varphi$ KK mode has a non-zero VEV, $u$, and $u \ll \pi/L$, then the 
top mass may be computed as in the $y_0=0$ case. The only notable difference
is that $m_t$ is enhanced by a factor of $\sqrt{F_1(0)/F_1(y_0)}$.
This factor reaches its maximum of $\sqrt{2}$ at $y_0 = L/2$.
It appears that the upper end of the interval for $n_{\rm KK}$
is preferred in this case.

In the more general case, where the VEVs of some $\varphi_j$ are comparable
with the compactification scale, one could imagine that the preferred
value of the string scale is lower, $M_s \sim 10$ TeV. 
In such a situation our estimates would no longer be reliable,
but the qualitative picture of a composite Higgs doublet bound out 
of $\psi_L$ and a tower of $t_R$ KK modes might remain valid.

Finally, we emphasize that the masses of the light quark and leptons
may easily be accommodated 
in our scenario. For example,
certain four-quark operators presumed to be generated at the string scale
with coefficients of order one in $M_s$ units, 
give rise in the low-energy effective theory to the Standard Model 
Yukawa couplings \cite{minimal}.

%%%%%%%%%%%%%%%%%%%%%%%%%%%%%%%%%%%%%%%%%%%%%%%%%%%%%%%%%%%%%%%%%%%%%%
%%%%%%%%%%%%%%%%%%%%% Section 5 %%%%%%%%%%%%%%%%%%%%%%%%%%%%%%%%%%%%%
%%%%%%%%%%%%%%%%%%%%%%%%%%%%%%%%%%%%%%%%%%%%%%%%%%%%%%%%%%%%%%%%%

\section{Conclusions}
\setcounter{equation}{0}

Electroweak symmetry breaking remains the foremost problem
facing elementary particle physics at this moment. We expect to
come to understand it in scientific detail in the next decade with the
Tevatron and the LHC.

We find it remarkable that the ingredients needed for a dynamical
explanation of the origin of the electroweak scale, which we often have
previously invoked in model building attempts
(e.g., topcolor, vector-like fermions,
strong coupling Nambu--Jona-Lasinio dynamics, etc.),
are seemingly presented automatically in
theories with extra-dimensions at the $\sim$ TeV scale.

In this paper we have explicitly constructed a ``demo--model''
of the dynamics in which the only fundamental
fields below the string scale are the $SU(3)_C \times SU(2)_W \times
U(1)_Y$
gauge bosons and the three generations of quarks and leptons, living in
a higher-dimensional compact space.

Strong dynamics comes from the existing QCD gauge group, which has a
large coupling strength above the compactification scale, due to the
large number of KK-modes. The KK-mode gluons act
like degenerate octets of colorons which, via exchange, give rise to
four-fermion operators.  Thus follows an NJL approximation
to the dynamics induced by these operators.

We find that various attractive channels lead to the formation of scalar
bound-states.
The Higgs doublet channel corresponds to $\bar{\chi} \psi_L $
where $\chi$ is the right-handed top quark field which we take to live
in the bulk. While $\chi$ has a chiral zero-mode by construction, which
is the $t_R$, the Higgs doublet emerges as a bound-state involving
a linear combination of the active KK-modes inherent in $\chi$.
In the effective
theory the large number of active
KK-modes, $n_{\rm KK}$, controls the dynamics, and naturally leads to a
tachyonic mass term for the Higgs at low energies, and thus electroweak
symmetry breaking.
We also expect various gauge-singlet composite bosons
to form in channels such as $\bar{\chi}\chi$, which somewhat complicate
the discussion of the low energy spectroscopy.  A low mass Higgs boson
may emerge through mixing between the primary composite Higgs and the
extra composite singlets.

Our model is largely intended to illustrate what can happen in the
extra-dimensional theories. It is hardly unique. The only
selection criterion seems to be the assignment of Standard Model fields
to the world-brane or into the bulk, in various
dimensional configurations.
We believe that, once the brane/bulk
field assignments are made in this manner, much of the
dynamics we describe is forced to happen.  New strong dynamics
is therefore natural and expected to occur in these theories.
The experimental confirmation of a strongly
interacting Higgs sector beyond the Standard Model would,
though not ``imply'', nonetheless lend support to
the notion of extra dimensions at the TeV scale.

\subsection*{Acknowledgements}
We would like to thank Bill Bardeen, Jeffrey Harvey, and Martin Schmaltz,
for valuable comments.
H.-C.~Cheng is supported by Department of Energy Grant DE-FG02-90ER-40560
and by a Robert R. McCormick Fellowship.
The research of B.~A.~Dobrescu is supported by NSF Grant PHY94-07194, and 
by DOE  Grant DE-AC02-76CH03000.

%%%%%%%%%%%%%%%%%%%%%%%%%%%%%%%%%%%%%%%%%%%%%%%%%%%%%%%%%%%%%%%%%%%%%%%%%%
%%%%%%%%%%%%%%%%%%%%% Section 5 %%%%%%%%%%%%%%%%%%%%%%%%%%%%%%%%%%%%%%%%%%%
%%%%%%%%%%%%%%%%%%%%%%%%%%%%%%%%%%%%%%%%%%%%%%%%%%%%%%%%%%%%%%%%%%%%%%%%

\section*{Appendix: effective potential parameters}
\renewcommand{\theequation}{A.\arabic{equation}}
\setcounter{equation}{0}

In this Appendix we give the formulae for the parameters of
the low-energy effective Lagrangian in the continuous approximation
by replacing sums of the KK states with the momentum integrals in
the fifth direction.

Cutting off the integrals at $\Lambda$ and replacing $\Lambda L/\pi$ by
$n_{\rm KK}$, we find the following wave function renormalizations
 at low-energy ($\sim L^{-1}$)
\bear
Z_H &\approx& n_{\rm KK} \frac{N_c c g_5^2}{16 \pi^2 L} \; 2\;
  F_1(y_0) ~, \nonumber \\ [2mm]
Z_\varphi &\approx& n_{\rm KK} \frac{N_c c g_5^2}{16 \pi^2 L} \;\frac{5}{4}
\;  F_2  ~.
\eear
Likewise, we find the parameters from the five-dimensional effective potential 
(see section 3.2):
\bear
\tilde{M}_H^2 &\approx& \Lambda^2 \left[ 1 - 
n_{\rm KK} \frac{N_c c g_5^2}{16 \pi^2 L}
  \; 4 \,F_3(y_0) \right]~,
\nonumber \\ [2mm]
\tilde{M}_\varphi^2 &\approx& \Lambda^2 \left[ 1 - 
n_{\rm KK} \frac{N_c c g_5^2}{16 \pi^2 L} \;\frac{5}{2} \,F_4  \right]~,
\nonumber \\ [3mm]
\tilde{\lambda}_H &\approx& n_{\rm KK}^2 \frac{N_c}{16 \pi^2}
\left(\frac{c g_5^2}{L}\right)^{\! 2}
   \;8 \;F_5(y_0) ~,
\nonumber \\ [2mm] 
\tilde{\lambda}_0 &\approx&  n_{\rm KK} \frac{N_c}{16 \pi^2}
\left(\frac{c g_5^2}{L}\right)^{\! 2}
   \; 5 \;F_6(y_0) ~,\nonumber \\ [2mm]
\tilde{\lambda}_\varphi &\approx&  n_{\rm KK} \frac{N_c}{16 \pi^2}
\left(\frac{c g_5^2}{L}\right)^{\! 2}
   \;\frac{75}{8} \;F_2 ~,
\eear
where the $F$-functions are defined by
\bear
F_1(y_0) &=& \int_0^1 p^2 dp^2 \int_0^1 dq \cos^2 (q \Lambda y_0)
   \frac{1}{p^2 (p^2+q^2)} \nonumber \\ [1mm]
F_2 &=& \int_0^1 p^2 dp^2 \int_0^1 dq
    \frac{1}{(p^2+q^2)^2} \nonumber \\ [1mm] 
F_3(y_0) &=& \int_0^1 p^2 dp^2 \int_0^1 dq \cos^2 (q \Lambda y_0)
   \frac{1}{p^2+q^2} \nonumber \\ [1mm]
F_4 &=& \int_0^1 p^2 dp^2 \int_0^1 dq
    \frac{1}{p^2+q^2} \nonumber \\ [1mm]
F_5(y_0) &=& \int_0^1 p^2 dp^2 \int_0^1 dq \cos^2 (q \Lambda y_0)
   \int_0^1 dq' \cos^2 (q' \Lambda y_0)
    \frac{1}{(p^2+q^2) (p^2+q'^2)} \nonumber \\ [1mm]
F_6(y_0) &=& \int_0^1 p^2 dp^2 \int_0^1 dq \cos^2 (q \Lambda y_0)
   \frac{1}{(p^2+q^2)^2} .
\eear
For $\psi_L$ localized at the boundary, we have
\bear
& & F_1(0) \, = \, \frac{\pi}{2} + \ln 2 \, \approx \, 2.26 ~, 
\nonumber \\ [2mm]
& & F_6(0) \, = \, F_2 \, = \,  \frac{\pi}{4}+\ln 2 \, \approx \, 1.48 ~, 
\nonumber \\ [2mm]
& & F_3(0) \, = \, F_4 \, = \, \frac{1}{3}+\frac{\pi}{6} - \frac{1}{3}\ln 2 \, 
\approx \, 0.63 ~, 
\nonumber \\ [2mm]
& & F_5(0) \, \approx \,  %3-\pi/4+0.50 
2.71 ~.
\eear
If $\psi_L$ is localized in the middle of the $[0,L]$ interval
and $\Lambda y_0 \gg 1$, the $\cos^2 (q \Lambda y_0)$ weigth factor 
averages to $1/2$, and therefore 
\bear
F_{1,3,6}\left(y_0 \sim L/2\right) & \approx & \frac{1}{2}F_{1,3,6}(0) ~, 
\nonumber \\ [1mm]
F_5\left(y_0 \sim L/2\right) & \approx & \frac{1}{4}F_5(0) ~.
\eear

%%%%%%%%%%%%%%%%%%%%%%%%%%%%%%%%%%%%%%%%%%%%%%%%%%%%%%%%%%%%%%%%
%%%%%%%%%%%%%%%%%%%%%%%%%%%%%%%%%%%%%%%%%%%%%%%%%%%%%%%%%%%%%%%

\vfil
\end{document}